\documentclass [11pt]{article} 
\usepackage{amsmath,amsthm,amsfonts,amscd,eucal,latexsym,latexsym,amssymb} 
\def\spa{\hskip -3pt}

\newsymbol\bt 1202           
\def\bC{{\mathbb C}}           

\def\bR{{\mathbb R}}

\def\bZ{{\mathbb Z}} 
 
\newsymbol\rest 1316         
 
\def\beq{\begin{eqnarray}}
\def\eeq{\end{eqnarray}}

\def\v{{\bf v}}
\def\m{{\bf m}}
\def\n{{\bf n}}
\def\S{{\bf S}}
\def\T{{\bf T}}
\def\K{{\bf K}}
\def\B{{\cal B}}

\begin{document} 
 
\hfill{\sl preprint - UTM 631} 
\par 
\bigskip 
\par 
\rm

 
\par 
\bigskip 
\LARGE 
\noindent 
{\bf  The interplay of the polar decomposition theorem and the Lorentz group} 
\bigskip 
\par 
\rm 
\normalsize 
 
 
 
\large 
\noindent {\bf Valter Moretti}

\noindent 
Department of Mathematics,  
 University of Trento and I.N.F.N. Gruppo Collegato di Trento,
via Sommarive 14,
I-38050 Povo (TN), Italy. \\
E-mail: moretti@science.unitn.it
\large 
\smallskip 
 
\rm\normalsize 
 

 
\par 
\bigskip 
\par 
\hfill{\sl November 2002} 
\par 
\medskip 
\par\rm

\noindent 
{\bf Abstract:} 
It is shown that the polar decomposition theorem of operators in (real) Hilbert spaces gives rise
to the known decomposition in boost and spatial rotation part of any matrix of
the orthochronous proper Lorentz group $SO(1,3)\spa\uparrow$. 
This result is not trivial because the polar decomposition theorem is referred 
to a positive defined  scalar product while the Lorentz-group decomposition theorem deals with 
the indefinite Lorentz metric.  A generalization  to infinite dimensional spaces can be given.
It is finally shown that the polar decomposition of $SL(2,\bC)$ is preserved by the covering 
homomorphism of $SL(2,\bC)$ onto $SO(1,3)\spa\uparrow$.

\section*{I. Introduction and notation.}
If $H$ is a, either real or complex, Hilbert space,
a bounded bijective operator $T:H\to H$  can be uniquely decomposed as both $T=UP$ and $T=P'U'$
where $U,U'$ are orthogonal/unitary operators and $P,P'$ are bounded self-adjoint positive operators.
These decompositions are called the {\em polar decompositions} of $T$.  
Consider the {\em special orthochronous Lorentz group} \cite{Ruhl,PCT,W}
\begin{eqnarray}
SO(1,3)\spa\uparrow \:\:\::= \{\Lambda \in M(4,\bR) \:\:|\:\: \Lambda \eta\Lambda^t = \eta\:,\: \mbox{det}\: \Lambda =1 \:,\:
\Lambda^0\:_0>0\}\label{so(1,3)} \:,\end{eqnarray}
where $M(n,\bR)$ denotes real vector space of real $n\times n$ matrices, $0$ in $\Lambda^0\:_0$ is referred to 
the first element of the canonical basis of $\bR^4$, $e_0,e_1,e_2,e_3$ and
 $\eta = \mbox{diag}(-1,1,1,1)$.  If $\Lambda \in SO(1,3)\spa\uparrow$ one may consider the 
polar decompositions $\Lambda = \Omega P=P'\Omega'$ where $\Omega,\Omega' \in O(4)$ and $P,P'$ are non singular, symmetric, positive 
matrices in $M(4,\bR)$.
{\em A priori} those decompositions could be physically meaningless because $\Omega$ and $P$ could not to belong 
to $SO(1,3)\spa\uparrow$: the notions of symmetry, positiveness, orthogonal group $O(4)$ are refereed to the positive 
scalar product of $\bR^4$ instead of  the indefinite Lorentz scalar product (similar comments can be made for the other
polar decomposition).
The main result presented in this work is that the polar decompositions of $\Lambda \in SO(1,3)\spa\uparrow$  are in fact
 physically meaningful.
Indeed, they coincides with the known {\em physical decompositions} of $\Lambda$  in spatial-rotation and boost parts (this fact 
also assures the uniqueness of the physical decompositions). In part, the result can be  
generalized to  infinite dimensional (real or complex) Hilbert spaces.
As a subsequent issue, considering the universal covering of $SO(1,3)\spa\uparrow$,
$SL(2,\bC)$ \cite{Ruhl,PCT,W}, we show that the covering homomorphism $\Pi : SL(2,\bC) \to SO(1,3)\spa\uparrow$
preserves the polar decompositions of $SL(2,\bC)$ transforming them into the analogous decompositions in
$SO(1,3)\spa\uparrow$.

\section*{II. Square roots and polar decomposition.}
 A {\em real} Hilbert space $H$ is a vector space equipped with a symmetric
 scalar product $(\cdot|\cdot)$ and complete with respect to the induced norm topology.
Henceforth we
 adopt the usual notation and definitions concerning adjoint, self-adjoint, unitary operators in 
 Hilbert spaces (e.g, see \cite{Rudin}), using them either in complex or real Hilbert spaces $H$.
 Moreover $\B(H)$ denotes the space of all bounded operators $T:H\to H$. $T\in B(H)$ is said {\em positive}
   ($T\geq 0$)
 if $(u|Tu)\geq 0$ for all $u\in H$.
 The lemma and the subsequent theorem below straightforwardly extend the {\em polar decomposition theorem}
 (Theorem 12.35 in \cite{Rudin}) encompassing both the real and the complex case.
  The proofs are supplied in the appendix. 
  (In complex Hilbert spaces  bounded positive operators are self-adjoint
\cite{Rudin}, in that case the self-adjointness property can be omitted 
 in the hypotheses and the thesis of the lemma and the theorem and elsewhere in this work.)\\

\noindent{\bf Lemma 1}. ({\bf Existence and uniqueness of (positive) square roots in Hilbert 
spaces}).
{\em Let $T\in \B(H)$ be a self-adjoint  positive operator where $H$ is a, either real or complex, Hilbert space.
There exists exactly one operator $\sqrt{T}\in \B(H)$ 
such that $\sqrt{T}^*=\sqrt{T}\geq 0$ and 
$\sqrt{T}^2 =T$. If $T$ is bijective, $\sqrt{T}$ is so.
$\sqrt{T}$ is said the {\bf (positive) square root} of $T$.}\\
 
\noindent{\bf Theorem 1}. ({\bf Polar Decomposition in either Real or Complex Hilbert spaces}).
 {\em If $T\in \B(H)$ is a bijective operator where $H$ is a, either real or complex, Hilbert space:\\ 
 {\bf (1)}  there is a unique decomposition $T= UP$,
 where $U$ is unitary, and $P$ is bounded, 
 bijective, self-adjoint and positive. In particular 
 $P = \sqrt{T^* T}$  and
 $U= T(\sqrt{T^* T})^{-1}$;\\
 {\bf (2)} there is a unique decomposition
 $T= P'U'$,
 where $U'$ is  unitary e  and $P'$ is bounded, bijective, self-adjoint and positive. In particular
 $U'=U$ and $P'= UPU^{*}$.}

\section*{III. Lorentz group and polar decomposition.}
Let us come to the main point by focusing attention on  the real Hilbert space $H=\bR^4$
endowed with the usual positive scalar product. In that case
$\B(H) = M(4,\bR)$. Unitary operators are orthogonal matrices, i.e., elements of of $O(4)$ and, if $A\in \B(H)$ 
the adjoint $A^*$ coincides with the transposed matrix $A^t$, therefore self-adjoint operators are symmetric matrices.
The Lie algebra of  $SO(1,4)\spa\uparrow$ (\ref{so(1,3)})
admits a well-known basis made of {\bf boost generators} $K_1,K_2,K_3$  
and  {\bf spatial rotation generators} $S_1,S_2,S_3$:
\begin{eqnarray}
\def\temp{\multicolumn{1}{c|}{0}}
\def\timp{\multicolumn{1}{c|}{1}}
\def\tomp{\multicolumn{1}{c|}{0}}
\def\tempd{\multicolumn{1}{c|}{0}}
\def\timpd{\multicolumn{1}{c|}{1}}
\def\tompd{\multicolumn{1}{c|}{0}} 
\def\tempt{\multicolumn{1}{c|}{0}}
\def\timpt{\multicolumn{1}{c|}{1}}
\def\tompt{\multicolumn{1}{c|}{0}}
 {K}_1 = \left[
\begin{array}{cccc}
\tomp & 1 & 0& 0\\ \cline{1-4}
\timp & 0 & 0& 0\\
\temp & 0 & 0&0 \\
\temp & 0 & 0& 0
\end{array}
\right]\:,\label{bK1}
\:\: {K}_2 = \left[
\begin{array}{cccc}
\tompd & 0 & 1& 0\\ \cline{1-4}
\tempd & 0 & 0& 0\\
\timpd & 0 & 0&0 \\
\tempd & 0 & 0& 0
\end{array}
\right]\:,\label{bK2}
\:\: {K}_3 = \left[
\begin{array}{cccc}
\tompt & 0 & 0& 1\\ \cline{1-4}
\tempt & 0 & 0& 0\\
\tempt & 0 & 0&0 \\
\timpt & 0 & 0& 0
\end{array}
\right]\:.\label{bK3}
\end{eqnarray}
\begin{eqnarray}
\def\temp{\multicolumn{1}{c|}{0}}
\def\timp{\multicolumn{1}{c|}{1}}
\def\tomp{\multicolumn{1}{c|}{0}}
\def\tempd{\multicolumn{1}{c|}{0}}
\def\timpd{\multicolumn{1}{c|}{1}}
\def\tompd{\multicolumn{1}{c|}{0}} 
\def\tempt{\multicolumn{1}{c|}{0}}
\def\timpt{\multicolumn{1}{c|}{1}}
\def\tompt{\multicolumn{1}{c|}{0}}
 {S}_i = \left[
\begin{array}{cccc}
\tomp & 0 & 0&  0\\ \cline{1-4}
\temp &   &  &  \\
\temp &  & T_i&  \\
\temp &  &  &  
\end{array}
\right]\label{Si} \:\:\:\mbox{with}\:\:\:
 {T}_1 = \left[
\begin{array}{ccc}
 0 & 0&  0\\
 0 & 0& \spa\spa-1 \\
 0 & 1&  0
\end{array}
\right]\:,\label{bS1}
\:\: {T}_2 = \left[
\begin{array}{ccc}
 0 & 0& 1\\
 0 & 0& 0 \\
  \spa\spa-1& 0& 0
\end{array}
\right]\:,\label{bS2}
\:\: {T}_3 = \left[
\begin{array}{ccc}
0 & \spa\spa-1& 0\\
1 & 0& 0 \\
0 & 0& 0
\end{array}
\right]\:.\label{bS3}
\end{eqnarray}
From now on $\K,\S,\T$ respectively denote the formal vector with components $K_1,K_2,K_3$,  
the formal vector with components $S_1,S_2,S_3$ and the formal vector 
with components $T_1,T_2,T_3$.
It is known (see the appendix) that the matrices $e^{\theta \n\cdot \T}$, 
$\theta\in \bR$ and $\n$ versor in $\bR^3$, are all of the elements of $SO(3)$. $\n$ is the rotation axis
with clockwise rotation angle $\theta$ of the rotation $e^{\theta \n\cdot \T}$. 
(The correspondence between pairs  $(\theta,\n)$ and  $SO(3)$ is one-to-one with the following exceptions:
$\theta =0$ individuates the trivial rotation $I$ for all $\n$, $(\theta,\n)$ and $(\theta +2k\pi,\n)$ with
$k\in \bZ$ individuate the same rotation and finally,
the pairs $(\pi,\n)$ and $(\pi,-\n)$ individuates the same rotation.)
The elements of one-parameter subgroups of $SO(1,3)\spa\uparrow$,
$\Lambda = e^{\theta \n\cdot \S}$, with $\theta\in \bR$ and $\n$ versor in $\bR^3$, do not affect the time coordinate 
of the two Minkowski coordinate systems related by $\Lambda$ and rotate the spatial axes by $e^{\theta \n\cdot \T}$. These elements 
are called {\bf spatial proper rotations}. They
give rise to a trivial faithful representation of $SO(3)$ in $SO(1,3)\spa\uparrow$. 
Conversely,  the (Lorentz) {\bf boosts} are the elements of one-parameter subgroups of $SO(1,3)\spa\uparrow$,
$\Lambda = e^{\chi \m\cdot \K}$ with $\chi \in \bR$ and $\m$ versor in $\bR^3$. 
(The correspondence between boosts and  pairs $(\chi,\n)$ is one-to-one 
with the following exceptions: $\chi=0$ defines the trivial  boost $I$ not depending on $\n$,
 $(\chi,\m)$ and $(-\chi, -\m)$ define the same boost.)
The vector $\v:= c(\sinh\chi) \m$ ($c>0$ being the velocity of light) has the components of the 
relative velocity of the two inertial frames  with Minkowski coordinate systems 
related by $\Lambda$.
The next theorem clarifies the interplay of boosts, spatial rotations and polar decomposition.\\

\noindent {\bf Theorem 2}. {\em If  $ UP = P'U 
=\Lambda$ (with $P'= UPU^t$)
are polar decompositions of $\Lambda \in SO(1,3)\spa\uparrow$:\\
{\bf (1)} $P,P',U\in SO(1,3)\spa\uparrow$, more precisely $P,P'$ are boosts and $U$ a spatial proper rotation;\\
{\bf (2)} there are no other decompositions of $\Lambda$ as a product of a Lorentz boost
and a spatial proper rotation different from the two polar decompositions above.}\\

\noindent{\em Proof}.
If $P\in M(4,\bR)$ we shall uses the representation:
\begin{eqnarray}
\def\temp{\multicolumn{1}{c|}{}}
\def\timp{\multicolumn{1}{c|}{C}}
\def\tomp{\multicolumn{1}{c|}{g}}
\def\tempd{\multicolumn{1}{c|}{0}}
\def\timpd{\multicolumn{1}{c|}{B^t}}
\def\tompd{\multicolumn{1}{c|}{0}} 
\def\tempt{\multicolumn{1}{c|}{0}}
\def\timpt{\multicolumn{1}{c|}{1}}
\def\tompt{\multicolumn{1}{c|}{0}}
 P = \left[
\begin{array}{cccc}
\tomp &  & B^t& \\ \cline{1-4}
\temp &   &  &  \\
\timp &   & A &  \\
\temp &   &  &  
\end{array}
\right]\:,\label{P}
\end{eqnarray}
where $g\in \bR$, $B,C\in \bR^3$ and $A\in M(3,\bR)$.\\
(1) We start by showing that $P,U\in O(1,3)$. As $P=P^t$, $\Lambda \eta\Lambda^t =\Lambda$ entails
$UP\eta PU^t = \eta$. As $U^t=U^{-1}$  and $\eta^{-1}=\eta$, the obtained identity is  equivalent to
$UP^{-1}\eta P^{-1}U^t = \eta$ which, together with 
$UP\eta PU^t = \eta$,
implies  $P\eta P= P^{-1}\eta P^{-1}$, namely $\eta P^2\eta = P^{-2}$,
where we have used $\eta=\eta^{-1}$ once again. Both sides are symmetric (notice that $\eta=\eta^t$) and positive
by construction, by Lemma 1 they admit unique  square roots which
must coincide. The square root of $P^{-2}$ is $P^{-1}$ while the 
square root of $\eta P^2\eta$ is $\eta P\eta$ since $\eta P\eta$ is symmetric positive and
$\eta P\eta\eta P\eta= \eta PP\eta = \eta P^2\eta$. We conclude that 
$P^{-1}= \eta P\eta$ and thus $\eta = P\eta P$ because $\eta=\eta^{-1}$. Since $P=P^t$ we have found that $P\in O(1,3)$
and thus $U = \Lambda P^{-1}\in O(1,3)$. Let us prove that $P,U\in SO(1,3)\spa\uparrow$.
$\eta = P\eta P^t$ entails $\mbox{det}\: P=\pm 1$,
on the other hand $P=P^t$ is positive and thus $\mbox{det}\: P \geq 0$ and $P^0\:_0 \geq 0$. As a consequence
  $\mbox{det}\: P =1$ and  $P^0\:_0 \geq 0$. We have found that $P\in SO(1,3)\spa \uparrow$. Let us determine 
  the form of $P$ using (\ref{P}).
$P=P^t$, $P\geq 0$ and $P\eta P= \eta$ give rise to
the following equations: $C=B$,
$0<g = \sqrt{1+B^2}$, $AB = gB$, $A=A^*$,  $A\geq 0$ and $A^2 = I+ BB^t$.
 Since $I+ BB^t$ is positive, the solution of the last equation $A= \sqrt{A^2} = I+ BB^t/(1+g) \geq 0$ is  the unique solution by Lemma 1.
We have found that a matrix $P\in O(1,3)$ with $P\geq 0$, $P=P^*$ 
 must have the form
 \begin{eqnarray}
\def\temp{\multicolumn{1}{c|}{}}
\def\timp{\multicolumn{1}{c|}{(\sinh \chi)\n}}
\def\tomp{\multicolumn{1}{c|}{\cosh \chi}}
\def\tempd{\multicolumn{1}{c|}{0}}
\def\timpd{\multicolumn{1}{c|}{(\sinh \chi)\n}}
\def\tompd{\multicolumn{1}{c|}{0}} 
\def\tempt{\multicolumn{1}{c|}{0}}
\def\timpt{\multicolumn{1}{c|}{1}}
\def\tompt{\multicolumn{1}{c|}{0}}
 P = \left[
\begin{array}{cccc}
\tomp &  & (\sinh \chi)\n^t& \\ \cline{1-4}
\temp &   &  &  \\
\timp &  I&\spa\spa\spa-\:(1-\cosh \chi){\n\n^t} &  \\
\temp &   &  &  
\end{array}
\right]\:,\label{P'}
\end{eqnarray}
where we have used the parameterization $B = (\sinh \chi) \n$, $\n$ being any versor in $\bR^3$ and $\chi\in \bR$.
 If $\n':= e^{\theta \m\cdot \T}\n$ (which is a versor since $e^{\theta \m\cdot \T}\n\in SO(3)$),
by direct computation 
it arises that:
\begin{eqnarray}
\def\temp{\multicolumn{1}{c|}{}}
\def\timp{\multicolumn{1}{c|}{(\sinh \chi)\n'}}
\def\tomp{\multicolumn{1}{c|}{\cosh\chi}}
\def\tempd{\multicolumn{1}{c|}{0}}
\def\timpd{\multicolumn{1}{c|}{B}}
\def\tompd{\multicolumn{1}{c|}{0}} 
\def\tempt{\multicolumn{1}{c|}{0}}
\def\timpt{\multicolumn{1}{c|}{1}}
\def\tompt{\multicolumn{1}{c|}{0}}
 e^{\theta \m\cdot \S} P \left(e^{\theta \m\cdot \S}\right)^t= \left[
\begin{array}{cccc}
\tomp &  & \sinh \chi(\n')^t& \\ \cline{1-4}
\temp &   &  &  \\
\timp &  I&\spa\spa\spa-\:(1-\cosh \chi){\n'{\n'}^t} &  \\
\temp &   &  &  
\end{array}
\right]\:,\label{RP'R}
\end{eqnarray}
It is simply proven that the matrix in the right hand side of (\ref{RP'R}) coincides with $e^{\chi \n'\cdot\K}$ if
$\n'= e_3$ and this happens for a suitable choice of parameters $\m_P,\theta_P$.  
Therefore we have the decomposition $P=  e^{\theta_P \m'\cdot \S} e^{\chi e_3 \cdot \K}\left(e^{\theta_P \m'\cdot \S}\right)^t$
for $\m'=-\m_P$. On the other hand, from the commutation relations $[S_i,K_j]= \sum_{k=1}^3\epsilon_{ijk} K_k$ it is simply proven that, for all versors $\m,\n$
 and $\theta\in \bR$:
$e^{\theta \m\cdot \S}\n\cdot \K \left(e^{\theta \m \cdot \S}\right)^t =  \left(e^{\theta \m \cdot \T} \n
 \right)\cdot \K$ 
 (the proof is based on the fact the functions of $\chi$ in both sides satisfy the same differential equation with the same 
 initial condition). As a consequence,
 \begin{eqnarray}e^{\theta \m \cdot \S} e^{\chi \n\cdot \K}\left(e^{\theta \m \cdot \S}\right)^t =
 e^{\chi \left(e^{\theta \m \cdot \T} \n
 \right)\cdot \K }\label{SK2}\:. \end{eqnarray} 
Specializing to the case $\n=e_3$, $\theta=\theta_P$ and $\m=\m'$, we have found that 
every matrix $P\in O(1,3)$ with $P\geq 0$ and $P=P^*$ can be written as $P= e^{\chi {\bf p} \cdot \K}$ for some
$\chi\in \bR$ and some ${\bf p}$ versor of $\bR^3$. In other words $P$ is a Lorentz boost. (The same proofs apply to $P'$.)\\
Let us pass to consider $U$. Since $\Lambda,P\in SO(1,3)\spa\uparrow$, from $\Lambda P^{-1}=U$ we conclude that $U\in SO(1,3)\spa\uparrow$. 
$U\eta =\eta (U^t)^{-1}$ (i.e. $U\in O(1,3)$)  and $U^t=U^{-1}$ (i.e. $U\in O(4)$) entail that $U\eta=\eta U$ and thus 
the eigenspaces of $\eta$, $E_\lambda$ (with eigenvalue $\lambda$), are invariant under the action of $U$. In those spaces $U$ acts as an element of 
$O(\mbox{dim}\:(E_\lambda))$ and the whole matrix $U$ has a block-diagonal form. 
$E_{\lambda=-1}$ is generated by $e_0$ 
and thus $U$ reduces to
$\pm I$ therein. The sign must be $+$ because of the requirement $U^0\:_0>0$. The eigenspace 
$E_{\lambda =1}$ is generated by $e_1,e_2,e_3$ and therein $U$ reduces to an element of $R\in O(3)$. Actually the requirement $\mbox{det}\: U=1$
 (together with $U^0\:_0 =1$) implies that $R\in SO(3)$ and thus $R=e^{\theta \m \cdot \T}$ for some versor $\m$ and some real $\theta$.
Using the found  block-diagonal  structure of the matrix $\Omega$ and the definition of the matrices $\S$ in functions of 
the matrices $\T$,
it is straightforwardly proven that  $\Omega = e^{\theta \m \cdot \S}$. \\
(2) If  $\Omega B=\Lambda \in SO(1,3)\spa \uparrow$ where $B$ is a pure boost and $\Omega$ is a spatial proper rotation. 
 $B = e^{\chi \n\cdot \K}$ is symmetric by construction since $K_i = K^t_i$. As a consequence of (\ref{SK2}) we find
 $e^{\chi \n\cdot \K} = (e^{\theta \m \cdot \S})^t e^{\chi K_3} e^{\theta \m \cdot \S}$  ($\m$ is 
 orthogonal to  $e_3$ and $\n$ and $\theta$ is the rotation angle around $\m$ of the rotation which transforms
  $e_3$ into $\n$). By direct inspection one see that $e^{\chi K_3}$ is positive and thus $B$ is so. On the other hand if $\Omega= e^{\theta \n \cdot \S}$, $\Omega^t= e^{\theta \n \cdot \S^t}=
 e^{-\theta \n \cdot \S} = \Omega^{-1}$ and thus $\Omega$ is orthogonal. We conclude that $\Lambda =\Omega B$
 is one of the two  polar decompositions (using the uniqueness property in Theorem 1). The proof for the other case $\Lambda = B'\Omega'$
 is strongly analogous. $\Box$\\

\noindent The result can be partially generalized into the following theorem. The proof is part of
the proof of the statement (1) of Theorem 2 with $\bR^4,\eta,\cdot^t$ replaced by $H,E,\cdot^*$ respectively.\\
 
 \noindent {\bf Theorem 3}.
  {\em Let $H$ be a, either real or complex, Hilbert space and
   $G_E$ the group of all of operators $\Lambda\in \B(H)$ 
 such that $\Lambda E \Lambda^* = E$, for a fixed
  $E\in \B(H)$ which is not necessarily positive and satisfies $E=E^{-1}=E^*$. The polar decompositions of $\Lambda \in G_E$,
$\Lambda= PU= UP'$ (where $U$ is the unitary operator) are such that $P,P',U\in G_E$ and the eigenspaces of $E$
are invariant for $U$.}\\

\noindent Notice that in the hypotheses above for $E$, $\sigma(E)\subset \{-1,+1\}$.\\ 

 \noindent Let us come to the last  result. As is well known, the simply connected Lie group $SL(2,\bC)$ is the universal covering of
  $SO(1,3)\spa\uparrow$ \cite{Ruhl,PCT,W}. Hence there is a surjective Lie-group homomorphism $\Pi: SL(2,\bC)\to SO(1,3)\spa\uparrow$
which is a local Lie-group isomorphism about each $L\in SL(2,\bC)$.\\

\noindent {\bf Theorem 4}.
{\em Let ${\bf \sigma}$ denote the vector whose components are the well-known {\em Pauli's matrices}
\begin{eqnarray}
 {\sigma}_1 = \left[
\begin{array}{cc}
  0 &1\\
  1& 0 
\end{array}
\right]\:,\label{si1}
\:\: {\sigma}_2 = \left[
\begin{array}{cc}
 0 & -i\\
 i & 0 
\end{array}
\right]\:,\label{si2}
\:\: {\sigma}_3 = \left[
\begin{array}{cc}
1 & 0 \\
0 & -1
\end{array}
\right]\:.\label{si3}
\end{eqnarray}
If $L\in SL(2,\bC)$ and $L=PU=UP'$ are its polar decompositions:\\
{\bf (1)} $P,P',U\in SL(2,\bC)$, in particular $P=e^{\chi \n\cdot {\bf \sigma}/2}$, $U=e^{-\theta \m\cdot i{\bf \sigma}/2}$
for some $\n,\m$ versors in $\bR^3$ and $\chi,\theta \in \bR$. \\ 
{\bf (2)} $\Pi(e^{\chi \n\cdot {\bf \sigma}/2}) =e^{\chi \n\cdot \K}$ and 
$\Pi(e^{-\theta \m\cdot i{\bf \sigma}/2}) = e^{-\theta \m\cdot \S}$ and thus $\Pi$
maps the polar decompositions of any $L\in SL(2,\bC)$ into the corresponding  polar decompositions of $\Pi(L)\in SO(1,3)\spa\uparrow$.}\\

\noindent{\em Proof}.
(1) We deal with the decomposition $PU$ only the other case being analogous.
As $0\leq P={P}^*$, $P$ can be reduced in diagonal form with positive eigenvalues so that $\mbox{det}\:{P}\geq 0$.
As a consequence $1=\mbox{det}\:{L}= \mbox{det}\:{P}\:\mbox{det}\:{U}$ entails that $\mbox{det}\:{U}> 0$. In turn,
the condition $U^{-1}=U^{*}$ implies $|\mbox{det}\:U|^2 =1$ and thus $\mbox{det}\:{U}=1$. We have proven that
$U\in SL(2,\bC)$ and also that and $P= LU^{-1}\in SL(2,\bC)$.
From the spectral theorem (see theorem 12.37 in \cite{Rudin}) there are two bounded self-adjoint 
operators $S,Q$ (i.e. Hermitean matrices of $M(2,\bC)$)
such that $P= e^{S}$ and $U=e^{iQ}$. Since the matrices $\sigma_):=I,\sigma_1,\sigma_2,\sigma_3$
are a basis of the real vector space of $2\times 2$ Hermitean matrices, $S= aI + \chi\n\cdot {\bf \sigma}$
and $Q= bI + \theta\m\cdot {\bf \sigma}$ for some versors $\n,\m\in \bR^3$ and reals $a,b,\chi,\theta$.
Using $\mbox{det}\:{e^X}= e^{\mbox{tr}\:X}$ and the fact that Pauli matrices are traceless,
the constraint $\mbox{det}\:{P}=\mbox{det}\:{U} =1$ implies $a=b=1$. This completes the proof of (1).\\
(2) By definition $\Pi$ maps a one-parameter subgroup with initial tangent vector $X$
into a one-parameter subgroup with initial tangent vector $d\Pi_IX$.
Since $\Pi(L)^i\:_j = \mbox{tr}\: \left( L \sigma_j L^*\sigma_i\right)/2$ 
where $i,j=0,1,2,3$ \cite{PCT}, it holds $d\Pi_I : -i\n\cdot{\bf \sigma}/2 \mapsto \n\cdot \S$ 
and similarly $d\Pi_I : \n\cdot{\bf \sigma}/2 \mapsto \n\cdot \K$
for $i=1,2,3$. Hence the one parameters groups $\theta \mapsto e^{-i\theta\m\cdot{\bf \sigma}/2}$ and $\chi \mapsto e^{\chi\n\cdot{\bf \sigma}/2}$
are respectively mapped into $\theta \mapsto e^{\theta\m\cdot\S}$ and $\chi \mapsto e^{\chi\n\cdot\K}$. $\Box$\\

\section*{A1. Proofs of some propositions.}
If $H$ is a real Hilbert space $H+iH$ denotes the complex Hilbert space
obtained by defining on $H\times H$: (i) the product $\:\:\:(a+ib) (u+iv) := au-bv + i(bu+av)\:\:\:$
where
$a+ib\in \bC$ and we have defined $u+iv:= (u,v)\in H\times H$, (ii)
the sum of $u+iv$ and $x+iy$ in $H\times H$: $\:\:\:(u+iv)+(x+iy) := (u+x) + i(v+y)\:\:\:$,
and (iii) the, anti-linear in the former entry, Hermitean scalar product $\:\:\:\langle u+iv| w+ix \rangle := (u|v)+ (v|x) +i(u|x) -i(v|w)$.
Let us introduce a pair of useful operators. The {\em complex conjugation} $\:\:J: u+iv \mapsto u-iv$
turns out to be an anti linear  operator with $\langle J(u+iv)| J(w+ix) \rangle 
= \langle w+ix| u+iv \rangle$ and  $JJ=I$.  
The unitary {\em flip operator} $C: u+iv \mapsto v-iu$ 
satisfies $C=C^*=C^{-1}$.
A bounded operator $A: H+iH \to H+iH$ is said to be {\em real} if $JA=AJ$. It is simply proven that,
(1) $A$ is real if and only if there is a (uniquely determined) 
pair of bounded operators  
$A_j:H\to H$, $j=1,2$, such that $\:\:\:A(u+iv) = A_1u + iA_2v\:\:\:$ for all $u+iv\in H+iH$;
(2) $A$ is real and  $AC=CA$, if and only if there is a (uniquely determined)  bounded 
operator  $A_0:H\to H$,  such that $\:\:\:A(u+iv) = A_0u + iA_0v\:\:\:$ for all $u+iv\in H+iH$.\\

\noindent {\em Proof of Lemma 1}. The proof in the complex case is that of  Theorem 12.33 in \cite{Rudin}.
Let us consider the case of a  real Hilbert space $H$.
If $T\in \B(H)$ is positive and self-adjoint, the operator on
$H+iH$, $A: u+iv \mapsto Tu + i Tv$ is bounded, positive and self-adjoint. By  Theorem 12.33 in \cite{Rudin}
there is only one $B\in \B(H+iH)$ with $0\leq B (= B^*)$ and $B^2=A$, that is the square root of $A$ which we 
indicate by  $\sqrt{A}$. 
Since $A$ commutes  with both $J$ and $C$, all of the
real polynomials in $A$  do so. 
If $\Omega\subset\sigma(A)$
is a Borel set and  $P_\Omega$ 
is the associated orthogonal projector in the spectral measure of $A$, 
there is a sequence of real polynomials in $A$ which tends 
to $P_\Omega$ in the strong operator topology (use Stone-Weierstrass' theorem and the fact that the space of 
continuous functions is dense in any $L^2(\bR,\mu)$ if $\mu$ is Borel with respect to the topology of $\bR$). Therefore every 
projector $P_\Omega$ commutes with both $J$ and $C$ and, in turn,
 every real Borel function of $A$ does so, $\sqrt{A}$  in particular.
We conclude that $\sqrt{A}$ is real with the form
$\sqrt{A} : u+iv \mapsto Ru +iRv$. The operator $\sqrt{T}:={R}$ fulfills all of requirements it being
bounded, self-adjoint and positive because $\sqrt{A}$ is so and
$R^2 = T$ since $(\sqrt{A})^2 = A: u+iv \mapsto Tu + i Tv$. If $T$ is bijective, $A$ is so by construction. Then, by Theorem 12.33 in \cite{Rudin}, $\sqrt{A}$ 
turns out to be bijective and, in turn, $R$ is  bijective too by construction.
Let us consider the uniqueness of the found square root. If $R'$ is another bounded
 positive self-adjoint square root of $T$, $B: u+iv \to R'u+iR'v$
is a bounded self-adjoint positive square root of $A$ and thus it must coincide with $\sqrt{A}$. This implies that $R=R'$.
 $\Box$\\

\noindent {\em Proof of Theorem 1}. 
(1) Consider the bijective operator $T:H\to H$ where $H$ is either real or
complex. $T^*T$ is bounded, self-adjoint, positive and bijective by construction.
Define $P := \sqrt{T^*T}$,
which exists and is bounded, self-adjoint, positive and bijective  by Lemma 1, and
 $U:= TP^{-1}$. $U$ is unitary because $\:\:\:U^*U= P^{-1}T^*TP^{-1}= P^{-1}P^2P^{-1}=I\:\:\:$,
where we have used $P^*=P$.
This proves that a polar decomposition of $T$ exists because $UP=T$ by construction.
Let us pass to prove the uniqueness of the decomposition.
If $T=U_1P_1$ is a other polar decomposition, 
$T^*T = P_1U^*_1U_1P_1 =  PU^*UP$. That is $P_1^2=P^2$.  Lemma 1 
implies that $P=P_1$ and  $U = T^{-1}P= T^{-1}P_1= U_1$.\\
(2) $P':=UPU^*$ is  bounded, self-adjoint, positive and bijective since $U^*$ is
unitary and $P'U'= UPU^*U=UP=T$. 
The uniqueness of the decomposition in (2) is equivalent to the
uniqueness of the polar decomposition $U'^*P'^*= T^*$ of $T^*$
which holds true by (1) replacing $T$ by $T^*$.
$\Box$\\

\noindent {\em Proof of the fact that $SO(3)$ is made by all of the matrices $e^{\theta \n\cdot T}$.} 
 If $R\in SO(3)$, the induced operator in  $\bR+i\bR$  is unitary and thus it admits a base of eigenvectors 
 with eigenvalues $\lambda_i$ with $|\lambda_i|=1$, $i=1,2,3$. As the characteristic 
polynomial of $R$ is real, an eigenvalue must be real, the remaining pair of eigenvalues being either real or complex and conjugates.
Since $\mbox{det} \:R= \lambda_1 \lambda_2 \lambda_3 =1$, $1$ is one of the eigenvalues. 
We conclude that $R$ has a real normalized eigenvector
$\n$ with eigenvalue $1$. By direct inspection one finds that  
$R$ is represented by the matrix $e^{\theta T_3}$ for some $\theta \in [0,2\pi]$
in any orthonormal base $\n_1:=\n$, $\n_2$, $\n_3$. In other words
$R= R'e^{\theta T_3} {R'}^t$ for some $R'\in SO(3)$.
 On the other hand $(T_i)_{jk}
=-\epsilon_{ijk}$ entails that $\sum_{i,j,k}U_{pi}U_{qj}U_{rk} \epsilon_{ijk} =\epsilon_{pqr}$ 
for all $U\in SL(3,\bR)$. That identity can be re-written as
$\n \cdot U\T U^{t} = (U\n) \cdot \T$ for every $U\in SL(3,\bR)$. By consequence,
if $U\in SO(3)$ it also holds $Ue^{\theta \n \cdot \T}U^t= 
e^{\theta (U\n) \cdot \T}$. Therefore, the identity found above for any $R\in SO(3)$,
$R= R'e^{\theta T_3} {R'}^t$ with $R'\in SO(3)$, can equivalently be written
as $R = e^{\theta \n \cdot \T}$
for some versor $\n= R'e_3$. Finally, every matrix $e^{\theta \n \cdot \T}$ belongs to $SO(3)$ because
$(e^{\theta \n \cdot \T})^t= e^{\theta \n \cdot \T^t}=e^{-\theta \n \cdot \T}=(e^{\theta \n \cdot \T})^{-1}$
and $\mbox{det}\:e^{\theta \n \cdot \T} = e^{\theta \n \cdot \mbox{tr}\:\T} = e^0=1$.

\end{document}